\documentstyle[11pt,twoside,jltp]{article}

\title{Ferromagnetic- and Superconducting-like Behavior of Graphite} 

\author{Y. Kopelevich$^{+}$, P. Esquinazi, J. H. S. Torres$^+$\address{Department 
of Superconductivity and
Magnetism, Institut f\"ur Experimentelle Physik II, Universit\"at Leipzig, Linn{\'e}str. 5, D-04103 Leipzig,
Germany}  and S. Moehlecke\address{Instituto de F{\'i}sica ``Gleb Wataghin", 
Universidade Estadual de Campinas, 
Unicamp 13083-970, Campinas, S\~{a}o Paulo, Brasil}}

\runninghead{Y. Kopelevich {\it et al.}}{Ferromagnetic and Superconducting Behavior}

\input psfig

\begin{document}

\begin{abstract}
We have identified ferromagnetic- and superconducting-like magnetization 
hysteresis loops in highly oriented pyrolytic graphite samples below and above room temperature. 
We also found that both behaviors are very sensitive to low-temperature -- 
as compared to the sample synthesis temperature -- heat treatment. The possible
contribution of magnetic impurities and why these do not appear to be the reason for the
observed phenomena is discussed.

\end{abstract}
\maketitle
\vspace{0.3in}

The discovery of superconductivity \cite{kra,ros}and ferromagnetism \cite{all} in C$_{60}$-based 
systems, as well as the discovery of carbon nanotubes \cite{iij}, i.e. graphite sheets wrapped into 
cylinders, has triggered a wide scientific interest on carbon-based materials. In particular, both 
ferromagnetic and superconducting correlations have been predicted for carbon 
nanotubes \cite{egg,kro,bal} as well as ferrimagnetism  in 
nanometer-size graphite fragments due to edge states \cite{nak,fuj,wak}. Recently, some of the authors 
of this paper have performed 
magnetoresistance measurements \cite{kop} on highly oriented pyrolytic graphite (HOPG). The
obtained results suggested the occurrence of zero-field as well as magnetic-field-induced superconducting 
correlations at temperatures below $T \sim ~$50~K. Motivated by these results and related works, 
we have studied carefully the magnetic properties of HOPG in the temperature range between 2 K
 and 800 K and in applied magnetic field oriented either parallel $(H \perp c)$ or perpendicular to the 
graphite basal planes $(H || c)$. Surprisingly, we have found that 
\begin{figure}
\centerline{\psfig{file=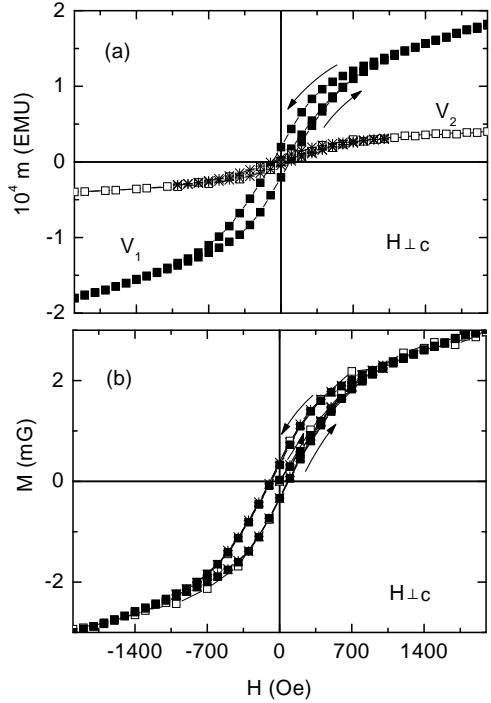,height=4.2in}}
\caption{(a) Magnetic moment $m(H)$ hysteresis loops measured in HOPG-2 samples with two different 
volumes ($V_1/V_2 \simeq 4.4)$ at $T = 100~$K and magnetic field applied parallel to the graphite basal 
planes $(H \perp c)$. We show for sample V$_2$ the hysteresis loops measured with two different field 
cycles $(\Box, \ast)$. (b) Magnetization $M(H) = m(H)/V$ data from (a) demonstrating that m(H) is proportional to the 
sample volume.} 
\label{1}
\end{figure}
HOPG shows both 
ferromagnetic- and superconducting-like 
magnetization hysteresis loops 
in a wide temperature range. The obtained results appear to rule out the possible influence of Fe-impurities. 
In this communication we
restrict ourselves to present the experimental evidence. Further results and characterization details
will be given in forthcoming publications.

The HOPG samples were prepared at the Research Institute ``Graphite" (Moscow) as 
described elsewhere \cite{bra} and characterized by means of x-ray diffraction, scanning electron 
microscopy (SEM), scanning tunneling microscopy (STM), spectrographic analysis and resistivity measurements. 
Spectrographic analysis indicate that the Fe concentration in our HOPG samples is $90 \pm 26~$ppm. 
X-ray diffraction $(\Theta-2\Theta)$ measurements give the crystal lattice parameters 
$a = 2.48~${\AA} and $c = 6.71~${\AA}. Both, x-ray rocking curves (FWHM $= 0.5^{\circ}$ to $1.4^{\circ}$) and SEM 
analysis show  a high degree of crystallites orientation along the hexagonal $c$-axis in the studied samples. 
The measured sample density is $2.26 \pm 0.01~$g/cm$^3$.
\begin{figure}
\centerline{\psfig{file=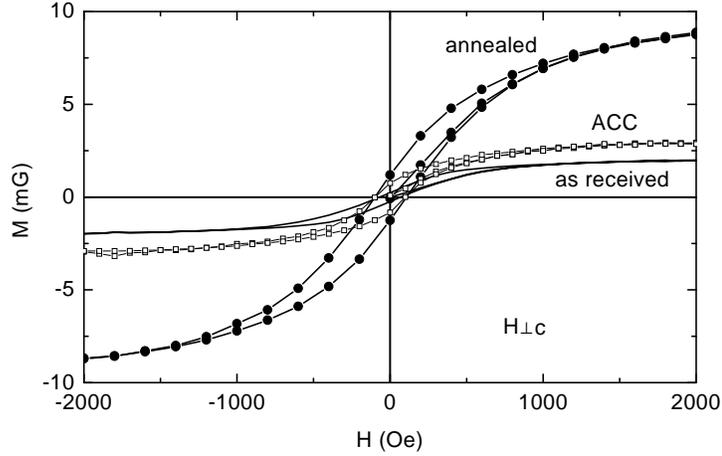,height=3.0in}}
\caption{Ferromagnetic-like $M(H \perp c)$ hysteresis loop measured at $T = 300~$K in the 
HOPG-2 sample ($V_3 =  2.45 \times 2.15 \times 1.9$~mm$^3$) before annealing (continuous line) 
at 800 K for
 2 hours, and $M(H \perp c)$ measured at $T = 350~$K after annealing ($\bullet$). 
$(\Box):$ $M(H \perp c)$ measured at $T = 300~$K on as-received sample from Advanced Ceramic Corporation 
(ACC) (the sample size is $3.6 \times 3.5 \times 1~$mm$^3$).}
\label{acc}
\end{figure}
The 
characteristic size of crystallites within the basal planes is about $1~\mu$m. Dc magnetization $M(H,T)$ 
measurements were performed with two SQUID magnetometers MPMS5 and MPMS7 from Quantum Design. 

\begin{figure}
\centerline{\psfig{file=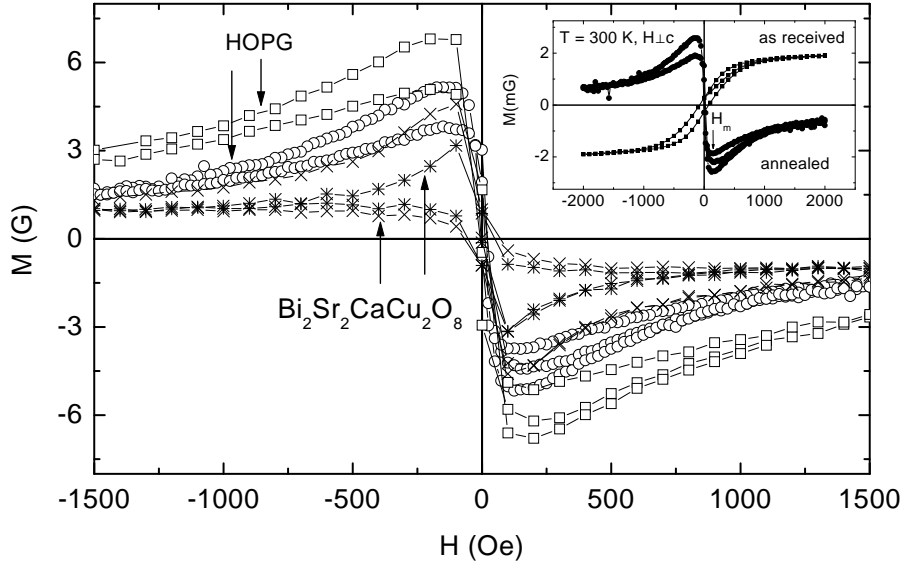,height=3.8in}}
\caption{Magnetization hysteresis loops $M (H \perp c)$ of annealed HOPG-1 sample ($V = 3.1 \times 1.75 \times 1.75~$mm$^3$) 
measured at $T = 300~$K $(\circ)$ and at $T = 150~$K $(\Boxð$). The magnetization of the HOPG-1 
sample was multiplied by a factor of 2000 for a better comparison to the $M(H||c)$ measured in a 
Bi2212 single crystal at $T = 45~$K $(\ast$) 
and $T = 35~$K $(\times)$. The inset shows $M(H \perp c)$ at $T = 300~$K measured 
before and immediately after annealing the graphite sample.}
\label{super}
\end{figure}

In the first part of this communication we present the data for field applied 
parallel to the basal planes 
($H \perp c$) where the diamagnetic magnetization\cite{her,dre} is small and no background subtraction is necessary
to identify the $M(H)$ hysteresis loops (Figs.~\ref{1},\ref{acc},\ref{super}). 
Figure \ref{1}(a) shows the ferromagnetic (FM)-like hysteresis loops of the magnetic moment 
$m(H)$ measured at $T = 100~$K and $H \perp c$ in two HOPG 
samples with dimensions 
$5 \times 4.8 \times 2.5~$mm$^3$ (V1) and $2.4 \times 2.3 \times 2.45~$mm$^3$ (V2) obtained
 from the same piece of graphite (labeled here as HOPG-2). Magnetization $M(H) = m(H)/V$ ($V$ is the 
sample volume) data are shown in Fig.~\ref{1}(b). As can be seen in Fig.~\ref{1}(a,b), $m(H)$ is 
proportional to the sample volume, demonstrating that the FM-like behavior is a bulk property of 
the sample. 
We note that $M(H)$ measured in the field intervals -10 kOe $~\le H \le~ +10$~kOe and 
-1 kOe~$\le H \le~ $+1 kOe are identical at low 
fields, so that the remnant magnetization at $H = 0$, $M_{\rm rem} = M_{hw}(0) = (M^+- M^-)/2$, 
 is the same for two different field cycles. 
Although the sample synthesis temperature is $\sim 2500~ ^{\circ}$C - $3000~ ^{\circ}$C, we 
 have found that a heat treatment at $T = 800~$K  under He exchange gas 
(the He pressure was  $\sim$10 Torr) during $\sim 2$~h can strongly enhance the ``ferromagnetism"
 for $H \perp c$,  see Fig.~\ref{acc}.

To illustrate how sensitive are the magnetic properties on the annealing details of HOPG we
show below the results obtained 
annealing the sample labeled HOPG-1 at $T = 800~$K during 2 hours 
in He exchange gas plus 2 hours in vacuum ($\sim 10^{-2}$ Torr). The measured $M(H)$-loops for $H \perp c$ 
 reveal that the FM-like 
behavior  of the as-received sample transforms to SC-like after annealing, as shown in the
inset of Fig.~\ref{super} for $T = 300~$K.
The similarity of the obtained $M(H)$ loops measured in the 
annealed graphite sample and those in superconductors is 
astonishing. As an example, we plot
in Fig.~\ref{super} the magnetization obtained at 
$T = 300~$K and at $T = 150~$K for graphite, as well as $M(H)$ measured in 
Bi$_2$Sr$_2$CaCu$_2$O$_8$ high-$T_c$ superconductor with $T_c = 89~$K at 
$T = 35~$K and $T = 45~$K.  Here we note only that the temperature dependence of the field corresponding 
to the minimum in $M(H)$ can be described by the equation 
\begin{figure}
\centerline{\psfig{file=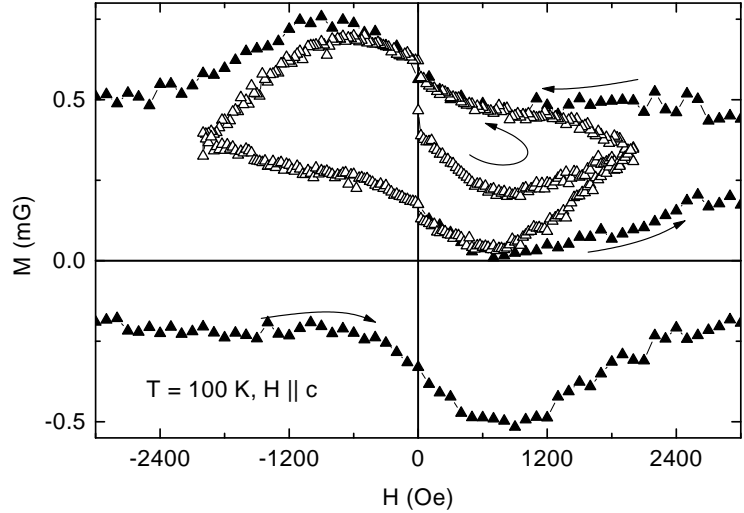,height=3.2in}}
\caption{Superconducting-like magnetization hysteresis loops $M(H)$ obtained in 
HOPG-2 sample (V$_1$) by cycling the magnetic field (applied parallel to the hexagonal $c-$axis)
 between -2 kOe and + 2 kOe (open triangles), and then  between -10 kOe and +10 kOe  (close triangles).
The curves were obtained subtracting a diamagnetic background signal $M = - \chi H$ with $\chi =  5.94 \times 10^{-2}$mG/Oe
from the measured signal.} 
\label{twoloops}
\end{figure}
$H_m(T) = H_0(1-T/T_0)$ with 
$H_0 \simeq 430~$Oe and $T_0 \simeq 500~$K. Interestingly,  the magnitude of $H_0$ is similar to 
that of the critical fields measured in superconducting graphite intercalation compounds \cite{dre}. 
It appears, however, that the SC-like hysteresis loops shown in Fig.~\ref{super} are metastable, such that
after one week the SC-like state relaxes toward the FM-like state.

For fields applied perpendicular to the basal planes, the major signal is due to their diamagnetic 
contribution. After subtraction of a linear diamagnetic background $M = - \chi(T) H$ we found SC-like hysteresis  loops
in the as-received samples.
Figure~\ref{twoloops}
shows the measured magnetization hysteresis loops 
obtained at $T = 100~$K for the sample V$_1$ after subtraction of a linear diamagnetic background signal.
\begin{figure}
\centerline{\psfig{file=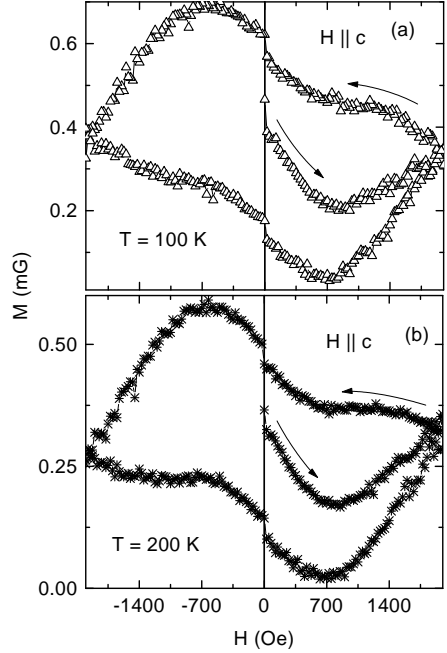,height=3.9in}}
\caption{Superconducting-like hysteresis loops $M(H || c)$ measured in HOPG-2 (sample V$_1$) for 
two temperatures demonstrating the occurrence of a zero-field magnetization jump. The
diamagnetic background signals $M = - \chi H$ were subtracted with: $\chi = 5.94 \times 10^{-2}~$mG/Oe in (a)
and $\chi = 4.938 \times 10^{-2}~$mG/Oe in (b).} 
\label{hopg2}
\end{figure}
The hysteresis loops resemble very much those known for superconductors. 
The loops presented in Fig.~\ref{twoloops} were measured in the field interval -2 kOe$~\le H \le~$ +2 kOe and 
-10 kOe$~\le H \le~$ +10 kOe (in Fig.~\ref{twoloops} we show only the low-field portion 
of this loop). Besides the SC-like behavior itself note also: (1) the pronounced 
dependence of the width of the hysteresis loop $(M_{hw})$ on the explored field interval,
(2) the occurrence of the spontaneous magnetization at zero applied magnetic field, and (3)
 the magnetization jump at $H = 0$. The minor hysteresis loop of Fig.~\ref{twoloops} is replotted in  
Fig.~\ref{hopg2}(a) to show clearly the discontinuity in $M(H)$ at $H = 0$. 
For completeness, the hysteresis loop and the zero-field magnetization jump measured at 
$T = 200$~K are also shown in Fig.~\ref{hopg2}(b).

The results described above indicate that the SC-like $M(H)$ hysteresis loops 
cannot be an artifact of a simple combination of the out-of-plane ``ferromagnetism" and 
the in-plane large diamagnetism. 
However, the data 
\begin{figure}
\centerline{\psfig{file=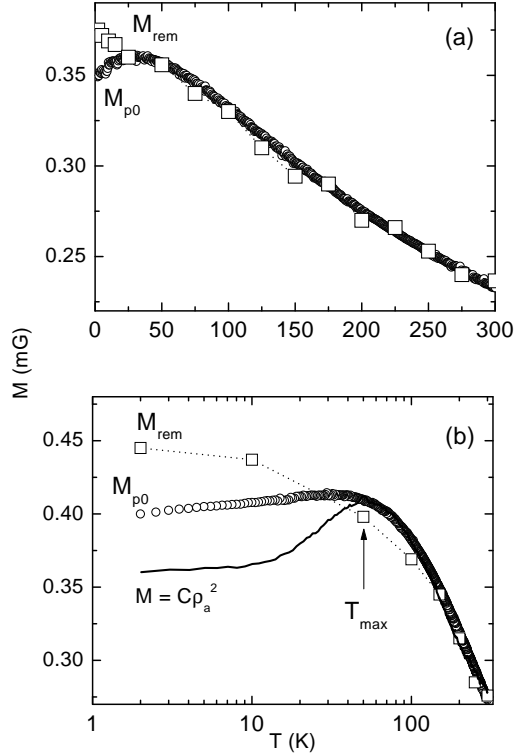,height=4.1in}}
\caption{(a) Temperature dependence of the remnant magnetization $M_{\rm rem} (\Box)$ and spontaneous 
magnetization $M_{p0} (\circ)$ (see text) measured in HOPG-2 (sample V$_1$); 
(b) $M_{\rm rem}$ and $M_{p0}$  measured in a similar sample (HOPG-1). The solid line has been obtained
by multiplying the square of the measured in-plane resistivity by a constant, i.e. $C \rho^2_a(T)$. $T_{\rm max}$
denotes the temperature where the resistivity shows a maximum.} 
\label{res}
\end{figure}
presented in Fig.~\ref{res} suggest that the SC- and FM-like phenomena 
may be coupled. In Fig.~\ref{res}(a) we show the remnant magnetization $M_{\rm rem}(T)$
obtained from the FM-like hysteresis loops $M(H \perp c)$ measured at various temperatures, as well as the 
magnetization $M_{p0}(T)$ (obtained at $H=0$ in the virgin state) 
related to the in-plane spontaneous currents measured
increasing temperature from 2 K to 300 K for sample V$_1$ (HOPG-2). We see that both 
quantities overlap at $T > 30~$K. Figure~\ref{res}(b) shows $M_{\rm rem}(T)$
 and $M_{p0}(T)$ measured in the sample HOPG-1, where in-plane resistivity 
measurements $\rho_a(T)$ have been also performed. As Fig.~\ref{res}(b) illustrates, $M_{p0}(T)$
 can be very well described by the equation $M_{p0} = C\rho_a^2(T)$ for $T > T_{\rm max}\sim 50~$K ($C$ is
a constant). 
Below $T \sim 30~$K$ < T_{\rm max}$ we observe that  $M_{p0}(T)$ starts to decrease with decreasing 
temperature as well. These results indicate a 
\begin{figure}
\centerline{\psfig{file=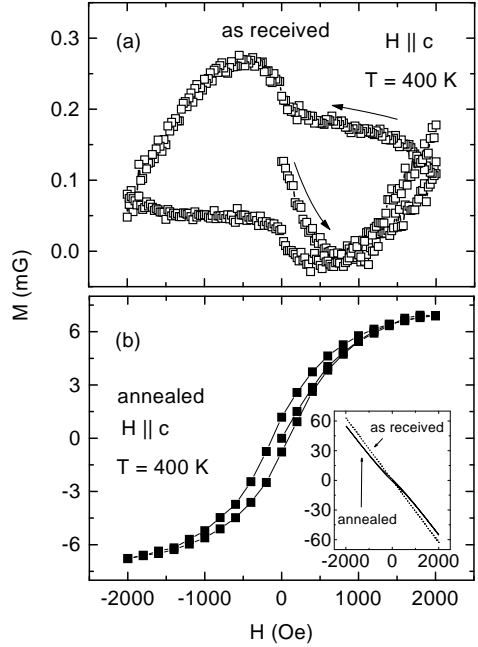,height=4.1in}}
\caption{The annealing-induced transformation of the SC-like magnetization $M(H || c)$ behavior
(a) to the FM-like one (b) obtained after annealing the as-received sample for 2 h in $\sim$10 Torr of He 
at 800~K. The diamagnetic background signals $M = - \chi H$ were subtracted:
$\chi = 3.147 \times 10^{-2}~$mG/Oe in (a) and $\chi = 3.1 \times 10^{-2}~$mG/Oe in (b).
The inset in (b) shows $M(H)$ without subtraction of the background
signals before and after annealing.}
\label{tran}
\end{figure}
close relationship between the magnetic,  resistive  and 
superconducting-like behavior in the HOPG sample. 

With annealing, the SC-like magnetic response measured in an as-received sample for $H || c$
 vanishes and FM-like  loops are obtained. We show in Fig.~\ref{tran} how a 
SC-like 
loop in $M(H || c)$ measured at $T = 400~$K in a virgin sample (a) transforms into the FM-like loop after annealing (b). 
It can be also seen in Fig.~\ref{tran}(a) that already at $T = 400~$K some irreversible changes in the sample properties 
take place. The inset in Fig.~\ref{tran}(b) presents the overall magnetization $M(H)$
 measured before and after annealing . As can be seen, the heat treatment does not change 
significantly the full diamagnetic signal.

We have also measured HOPG obtained from Union Carbide and Advanced Ceramics 
Corporation (ACC). In these samples the FM-like hysteresis 
\begin{figure}
\centerline{\psfig{file=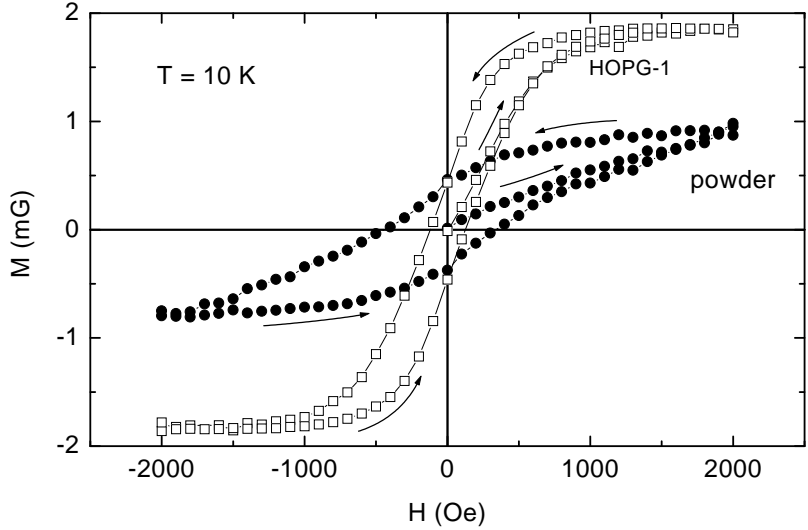,height=3.5in}}
\caption{Magnetization loops for the HOPG-1 sample for $H \perp c ~(\Box)$  
obtained without subtraction of a diamagnetic background signal, and for
the ultra pure spectroscopic graphite powder ($m = 98.7~$mg, $(\bullet)$) after
subtraction of a diamagnetic background signal $M = -\chi H$ with $\chi = 4.7 \times 10^{-2}~$mG/Oe.}
\label{powder}
\end{figure}
loops were measured for both field
orientations. The $M(H \perp c)$-loop measured at $T = 300~$K in the HOPG from ACC is shown in Fig.~\ref{acc}. 

Since the here studied HOPG-samples contain about 90 ppm Fe impurities, their possible effect should
be discussed. Note first that the 90 ppm Fe, if ferromagnetic, would induce a magnetization
 two orders of magnitude larger than the one found experimentally. On the other hand, formation
of ferromagnetic magnetite Fe$_3$O$_4$ clusters combined with the
antiferromagnetic 
Fe$_2$O$_3$ inclusions cannot be excluded. 
Assuming that this last compound may transform with annealing in magnetite - a rather questionable transformation 
taking into account our annealing  conditions - we would expect an enhancement of  the FM moment.
However, the following facts rule out such an interpretation. (a) 
With annealing, we observed both a strong enhancement of a FM-like behavior (see Fig.~\ref{acc}) 
and a transformation of FM-like $M(H \perp c)$ hysteresis loops to  SC-like ones (see Fig.~\ref{super}),
obtained without any background signal
subtraction. (b) For all studied HOPG samples, no anomalies in $M(H,T)$ were detected 
at $T \sim 120~$K (the Verwey transition temperature of Fe$_3$O$_4$) and at $T \sim 260~$K (the
crossover temperature from antiferro- to weak ferromagnetism in Fe$_2$O$_3$), see Fig.~\ref{res}.
(c) In the vicinity of the Curie temperature ($T_C = 860~$K for Fe$_3$O$_4$) a rapid rise of the
saturated magnetization must be observed. In contrast, after annealing we found a nearly temperature
independent magnetization $M(H \perp c)$ in the temperature range 700~K~$\le T \le 800~$K.
(d) While the remnant magnetization is independent on the cycling field amplitude for $H \perp c$
(see Fig.~\ref{1}), it is strongly dependent for $H || c$, see Fig.~\ref{twoloops}.
(e) The presence of FM clusters cannot account for the zero-field magnetization jump (Fig.~\ref{hopg2})
observed at all measured temperatures and only for $H || c$ orientation.
(f) The $M_{p0}(T)$ dependence, see Fig.~\ref{res}, demonstrates a positive curvature at high enough
temperatures and a maximum at $T \sim 30~$K. Both results appear to be inconsistent with a
spontaneous magnetization behavior in ferromagnets. 
(g) To check further a possible contribution of magnetic impurities, we have measured
an ultra-clean spectroscopic graphite powder with a characteristic particle size of
$\sim 50~\mu$m \cite{hoh}. Figure \ref{powder} shows the hysteresis loops for sample
HOPG-1 ($H \perp c$) and the one corresponding to the graphite powder at $T = 10~$K.
In spite of 4 orders of magnitude less Fe-concentration in the graphite powder, we measure
a similar remnant magnetization for both samples. Therefore, our studies indicate that the 
``ferromagnetism" in graphite appears to be an intrinsic property, 
not appreciated so far.

The interplay between SC- and FM-like behavior may not be surprising if we take into account
that all contributions to the magnetization of graphite come from carrier states situated 
in the vicinity of the Fermi level \cite{sha}. While at this stage of the research it is too 
premature to speculate
on the underlying mechanisms for the surprising phenomena observed, we would like
to include a few comments. 

Assuming 
that the SC-like behavior of HOPG is indeed associated with high-temperature superconductivity
which develops on a local scale (the resistivity of the bulk samples does not
go to zero \cite{tmax}), we tend to relate the occurrence of the  
magnetization $M_{p0}(T)$ to spontaneous currents induced at normal metal 
- superconductor interfaces 
recently discussed in the literature \cite{fau}. Coupled with this phenomenon, a first order 
transition\cite{fau}  at 
$H = 0$ then may explain the zero-field magnetization jump (see Fig.~\ref{hopg2}). Also, a first order transition 
implies a complex hysteretic behavior \cite{set,viv}. In principle, superconductivity can be localized at the intergrain 
boundaries of the crystallites in graphite. The edge states at the boundaries \cite{nak,fuj,wak}can be also responsible for 
the ferromagnetism. We note that the STM analysis has revealed steps of $6 \pm 1~$ {\AA}  height at the intergrain boundaries 
which may be considered as the edges of graphite fragments. Finally, we note that the extreme 
sensitivity of the magnetic behavior to heat treatment as well as the observed aging effect 
(time-induced transition from SC-like to FM-like behavior) can be related to the 
``low-temperature" kinetic \cite{zha}of graphitic structure at the edges. 

In summary, our studies reveal  high temperature ferromagnetic-like behavior and possible ``hot" 
superconductivity in highly oriented pyrolitic graphite. It appears to be
difficult to interpret the observed phenomena as due to magnetic impurities. 

\section*{ACKNOWLEDGMENTS}
We thank R. H\"ohne and  M. Ziese for discussions and for the interest on this work, 
A. Setzer and J. Lenzner for technical assistance, B. Kohlstrunk and M. L\"osche for the
STM characterization, H. Semmelhack for the x-ray measurements, J. Becker  and  H.-J. Dietze 
(J\"ulich) for the
spectographic analysis.
We thank A. S. Kotosonov (Graphite Institute) for providing us with the HOPG samples and
V. V. Lemanov (St. Petersburg) for the collaboration. 
This work was partially supported by the
Innovationskolleg 	
``Ph\"anomene an den Miniaturisierungsgrenzen" (DFG IK 24/B1-1, project H), 
DAAD No. 415-bra-probral/bu, CAPES proc. No. 077/99, No. DS-44/97-0, CNPq proc. No. 301216/93-2,
No. 300862/85-7, and FAPESP proc. No. 95/4721-4, No. 99/0779-9.

\end{document}